\documentclass[conference]{IEEEtran}
\usepackage[utf8]{inputenc}
\usepackage[cmex10]{amsmath}
\usepackage{upgreek}
\usepackage{booktabs}
\usepackage{epsfig}
\usepackage{latexsym}
\usepackage{multirow}
\usepackage{stfloats}
\usepackage{epstopdf}
\usepackage{color}  
\usepackage{tabularx} 
\usepackage{algorithm}
\usepackage{algpseudocode}
\usepackage{amssymb}
\usepackage{enumerate}
\usepackage{array}
\graphicspath{{./Figures/}}
\usepackage{color}
\usepackage{bbm}
\usepackage{bm}
\usepackage{cite}
\usepackage[tight,footnotesize]{subfigure}
\usepackage{balance}
\usepackage{mathrsfs}
\usepackage{verbatim}
\usepackage{dsfont}
\usepackage{verbatim}
\usepackage{tikz}
\usepackage{setspace}
\usepackage{diagbox}
\usepackage{multicol}
\usepackage{environ}
\usepackage{tikz}
\usepackage{amsmath}
\usepackage{stfloats}
\usepackage{algorithm}
\usepackage{algpseudocode}
\usepackage{amsmath}
\usepackage{graphics}
\usepackage{epsfig}
\usepackage{caption}
\allowdisplaybreaks[4]
\usepackage{amsmath}
\usepackage{amsthm}
\usepackage{authblk}
\newcounter{TempEqCnt}

\def\BibTeX{{\rm B\kern-.05em{\sc i\kern-.025em b}\kern-.08em
		T\kern-.1667em\lower.7ex\hbox{E}\kern-.125emX}}
\begin{document}
\title{Scintillation and Attenuation Modelling of Atmospheric Turbulence for Terahertz UAV Channels
}
\author[1]{Weijun Gao}
\author[1]{Chong Han}
\author[2]{Zhi Chen}
\affil[1]{Shanghai Jiao Tong University, China. Email:  \{gaoweijun, chong.han\}@sjtu.edu.cn}
\affil[2]{University of Electronic Science and Technology of China, China. Email: chenzhi@uestc.edu.cn}
\def\allfiles{}
	\maketitle
	\thispagestyle{empty}
	\begin{abstract}
        Terahertz (THz) wireless communications have the potential to realize ultra-high-speed and secure data transfer with miniaturized devices for unmanned aerial vehicle (UAV) communications. Existing THz channel models for aerial scenarios assume a homogeneous medium along the line-of-sight propagation path. However, the atmospheric turbulence due to random airflow leads to temporal and spatial inhomogeneity of the communication medium, motivating analysis and modelling of the THz UAV communication channel. In this paper, we statistically modelled the scintillation and attenuation effect of turbulence on THz UAV channels. Specifically, the frequency- and altitude-dependency of the refractive index structure constant, as a critical statistical parameter characterizing the intensity of turbulence, is first investigated. Then, the scintillation characteristic and attenuation of the THz communications caused by atmospheric turbulence are modelled, where the scintillation effect is modelled by a Gamma-Gamma distribution, and the turbulence attenuation as a function of altitude and frequency is derived. Numerical simulations on the refractive index structure constant, scintillation, and attenuation in the THz band are presented to quantitatively analyze the influence of turbulence for the THz UAV channels. It is discovered that THz turbulence can lead to at most $10~\textrm{dB}$ attenuation with frequency less than $1~\textrm{THz}$ and distance less than $10~\textrm{km}$.
	\end{abstract}
	
	\section{Introduction}
	With the increasing demand for faster and more secure wireless communications, Terahertz ($0.1-10~\textrm{THz}$) communications have attracted great research attention for 6G and beyond wireless communication networks~\cite{han2022terahertz}. Thanks to the sub-millimeter wavelength and multi-tens-of-GHz continuous bandwidth, the THz band can support fast, highly directional, and secure wireless links. These unique spectrum features are beneficial to wireless communications among unmanned aerial vehicles (UAVs). First, aerial communication applications such as UAV-aided ubiquitous coverage and UAV-based relaying rely on transmissions of high-resolution images and low-latency commands, which requires ultra-high-speed data transmission. Second, miniaturized communication devices can lighten the load of small UAVs. Finally, the high directivity and the enhanced communication security can effectively prevent the UAVs from being eavesdropped through the aerial channels with bare obstructions.
	
	Modeling of the THz UAV channel model is one of the most fundamental topics in studying THz wireless communications between flying UAVs. Due to the lack of multi-paths in aerial communication scenarios, the THz UAV channel is assumed to transmit through the line-of-sight (LoS) path between the transceivers. In prior studies~\cite{li2021ray}, the LoS signal attenuation can be modelled by the summation of the free-space path loss, the molecular absorption effect led by molecules mainly consisting of water vapor, and the scattering effect due to small particles like raindrops, dust, and snowflakes under extreme weather conditions. These channel models assume that the propagation medium is homogeneous while for aerial communication scenarios, atmospheric turbulence led by wind can randomly change the temperature, pressure, and molecule component of the propagation medium. These lead to inhomogeneity of the medium and makes the existing models based on homogeneous medium invalid. Some experimental research studies such as~\cite{ma2015experimental} have discovered that turbulence can impair the data rate of THz communications and its influence is essential to be investigated.

    For the aforementioned motivations, it is necessary to analyze and model the effect of atmospheric turbulence in THz UAV communications.    
    Since the turbulence is deterministically governed by the Navier-Stokes equations~\cite{temam2001navier}, the modeling of scintillation and attenuation caused by turbulence meets mathematical difficulty in solving this non-linear equation. As a result, only statistical models are accessible to characterize the turbulence.
    Early statistical studies on the influence of turbulence mainly focus on the visible light frequency band for free-space optical communications rather than THz communications~\cite{esmail2021experimental}.
    
    In this paper, we statistically model the scintillation and attenuation effect caused by atmospheric turbulence for THz UAV channels. Specifically, we first analyze the refractive index structure constant (RISC), which is a key parameter to characterize turbulence. We develop the model of RISC at different frequencies and altitudes in the THz band based on the statistical turbulence model in the visible light frequency band. Second, we model the scintillation caused by turbulence by using a Gamma-Gamma distribution, which is a universal model applicable to the turbulence of various intensities. Finally, the scintillation and attenuation caused by turbulence at different propagation distances and RISC are evaluated to quantitatively demonstrate the influence of turbulence on THz UAV channels. 

	The remainder of the paper is organized as follows. In Sec.~\ref{sec:sys}, the THz LoS channel models in the homogeneous and inhomogeneous media are described and modelled, where the molecular absorption and scattering effect are presented. Particularly in inhomogeneous medium, the atmospheric turbulence in the THz band is statistically characterized. In Sec.~\ref{sec:scintillation_and_attenuation}, the scintillation and attenuation effects of turbulence on the THz UAV communications are investigated and computed. Numerical results for the turbulence including the altitude-dependent refractive index structure constant, Rytov variance, scintillation, and attenuation are evaluated in Sec.~\ref{sec:numerical} to quantitatively measure the influence of atmospheric turbulence on THz communications. The paper is concluded in Sec.~\ref{sec:concl}.

	\section{System Model}~\label{sec:sys}
	\begin{figure}
		\centering
        \includegraphics[width=\textwidth/2]{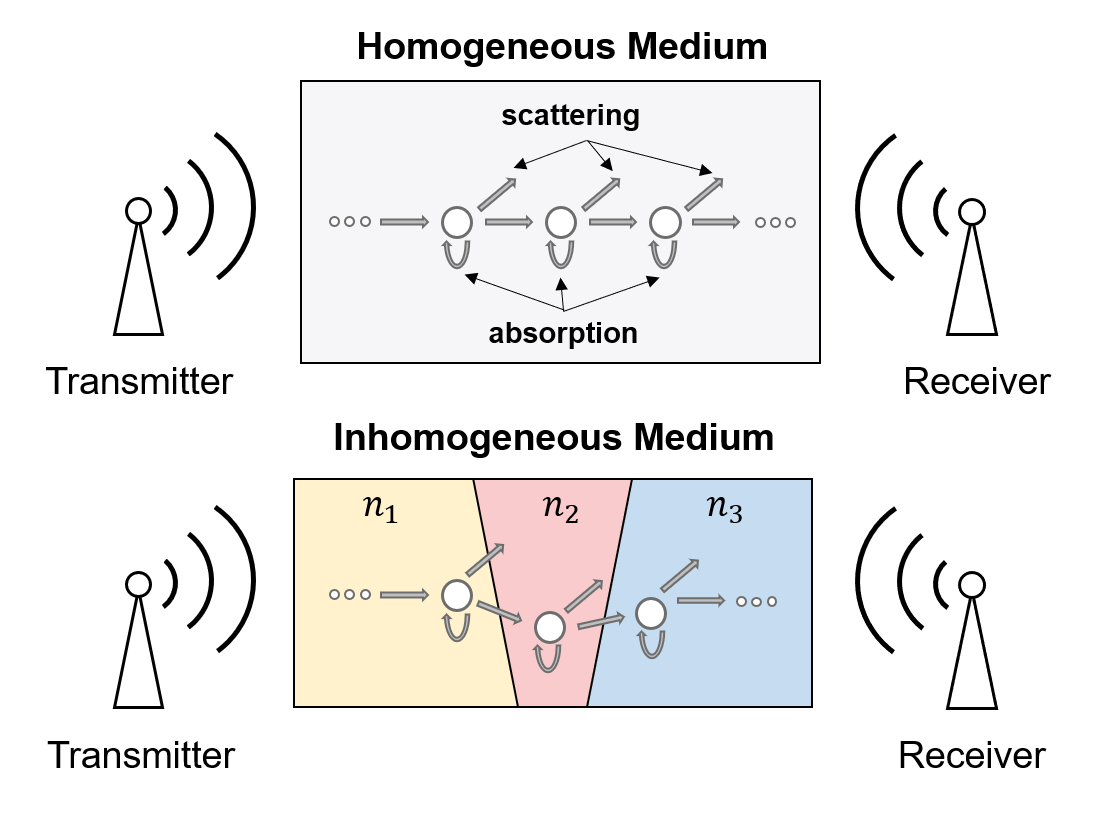}
		\captionsetup{font={footnotesize}}
    \caption{Illustration of Terahertz wave propagation in homogeneous and inhomogeneous mediums.}
	    \label{fig:system_model}
	\end{figure}
	We consider a THz UAV point-to-point wireless link as shown in Fig.~\ref{fig:system_model}. Since there are few obstacles in the aerial communication scenarios and THz wireless communications are highly directional, we assume that the LoS path between the transmitter and the receiver is not blocked and no multi-path effect is considered.  
	Besides free-space spreading loss, THz LoS signals experience absorption and scattering loss due to the interaction between the electromagnetic wave and the molecules and small particles in the medium. On one hand, molecules such as water vapor and oxygen can be excited by the THz wave and absorb part of the energy of the EM wave, which is referred to as the molecular absorption effect~\cite{jornet2011channel}.
	On the other hand, the small particles caused by extreme weather like rain, snow, dust, and fog can scatter the EM wave and lead to additional scattering loss.
	The signal attenuation caused by these two effects has been well studied in many previous studies on THz communications~\cite{jornet2011channel, ITUR}, which assume that the propagation medium is homogeneous along the propagation medium.
	However, in UAV with airflow, the THz propagation medium is not ideally homogeneous along the signal transmission path. At different altitudes, environmental parameters such as temperature, pressure, and moisture are different, which leads that the cross-altitude THz wave propagation having a spatially inhomogeneous medium. 
	The inhomogeneity is mainly caused by turbulence, which is caused by random airflow in the atmosphere. Unlike the homogeneous medium, the refractive index of the inhomogeneous medium is not uniform and the transmitted LoS signal can thus be randomly distorted.
	\subsection{Terahertz Line-of-Sight Wave Propagation in Homogeneous Medium}
	Since the component of air varies slowly with altitude, the propagation medium can be viewed as homogeneous for THz links with short vertical distances. THz wave propagation in a homogeneous medium experiences free-space path loss, molecular absorption loss, and scattering loss. 
	The LoS channel impulse response due to absorption and scattering can be respectively expressed by 
	\begin{align}
	    L_{\textrm{abs}} &= e^{k_{\textrm{abs}}(f,h)L},\\
	    L_{\textrm{sca}} &= e^{k_{\textrm{sca}}(f)L},
	\end{align}
	where $k_{\textrm{abs}}$ and $k_{\textrm{sca}}$ stand for the attenuation coefficient caused by the molecular absorption effect and the scattering effect, respectively.	The molecular absorption coefficient $k_{\textrm{abs}}$ is characterized in~\cite{jornet2011channel}, which is a frequency- and altitude-dependent. $f$ stands for the frequency and $h$ is the altitude. $L$ denotes the propagation distance. Water vapor dominates the molecular absorption with six orders of magnitude higher than oxygen and others, and thus we can express $k_{\textrm{abs}}(f,h)$ as
	\begin{equation}
	\begin{aligned}
	    k_{\textrm{abs}}(f,h)\approx k_{H_2O,\textrm{grd}}(f)\alpha_{H_2O}(h),
	\end{aligned}
	\end{equation}
	where $k_{H_2O,\textrm{grd}}(f)$ denotes the terrestrial water vapor absorption coefficient and $\alpha_{H_2O}(h)$ represents the ratio of water vapor density at altitude $h$ to the terrestrial one.
	
	The scattering effect between the EM wave and a certain type of particle can be classified into two cases, namely Rayleigh scattering and Mie scattering, depending on the relationship between the wavelength and the size of the particle. If the wavelength is larger than the size of the particle, the scattering is Rayleigh scattering, or otherwise, it becomes Mie scattering. At THz frequencies, the wavelength at millimeters or sub-millimeters is smaller than the radium of common particles such as rain, fog, and snow on the order of $10-100~\textrm{mm}$~\cite{ma2018effect}, and therefore the scattering in the THz band is mainly Mie scattering. The Mie scattering loss coefficient $k_{\textrm{sca}}$ in dB/km can be represented by
	\begin{equation}
	    k_{\textrm{sca}}[\textrm{dB/km}]=4.343\int_{0}^{\infty}\sigma_{\textrm{ext}}(r)N(r)dr,
	\end{equation}
	where $r$ denotes the radius of the particle, and $N(r)$ represents the molecular size distribution commonly modelled by an exponential distribution $N(r)=N_0\exp(-\rho_0 r)$. $\sigma_{\textrm{ext}}(r)$ stands for the extinction cross-section in Mie theory~\cite{norouzian2019rain}, which can be expressed by
	\begin{equation}
	\begin{aligned}
        \sigma_{\textrm{ext}}
        &=\frac{2\pi}{\chi^2}\sum_{m=1}^{\infty}(2m+1)\textrm{Re}(a_m+b_m)\\
        &\approx\frac{2\pi}{\chi^2}\sum_{m=1}^{x+4x^{1/3}+2}(2m+1)\textrm{Re}(a_m+b_m),
	\end{aligned}
	\end{equation}
	where $\chi=\frac{2\pi}{\lambda}$ is the wave number, and the threshold parameter is given by $x=\frac{2\pi r}{\lambda}$. $\textrm{Re}(\cdot)$ returns the real part of a complex number.
	$a_m$ and $b_m$ represent Mie scattering coefficients~\cite{norouzian2019rain}.
	In summary, the total path loss of the THz LoS channel in a homogeneous medium including free-space path loss, molecular absorption effect, and scattering can be expressed by
	\begin{equation}
	    L_{\textrm{tot}}^{\textrm{hom}}[\textrm{dB}]=20\log_{10}\left(\frac{4\pi fL}{c}\right)+0.434(k_{\textrm{abs}}+k_{\textrm{sca}})L,
	\end{equation}
	where $c$ stands for the speed of light.
    \subsection{Terahertz Line-of-Sight Wave Propagation in Atmospheric Turbulence}
    Atmospheric turbulence caused by airflow leads to random fluctuations of temperature, pressure, and water vapor density in propagation medium in time and space. 
    These fluctuations lead to a random fluctuation of the refractive index of the medium, which randomly distorts the EM wave. 
    Characterizing the turbulent flow is critical in analyzing its influence on THz wave propagation.
    Navier-Stokes equations, as fundamental to describe the motion of viscous fluid in hydrodynamics, are the most straightforward way to model atmospheric turbulence. However, due to the difficulty in mathematically solving those nonlinear equations, it is impractical to directly solve them to model the effect of turbulence on wave propagation. Moreover, different from how the attenuation and scattering in the homogeneous medium are characterized, it is difficult to precisely acquire the refractive index along the propagation path in the turbulent flow at any time or at any position. Therefore, only statistical models are feasible for THz wave propagation analysis in a turbulent flow.
     
     As the most fundamental statistical theory, Kolmogorov's theory is widely used to model the turbulence~\cite{kraichnan1964kolmogorov}, where turbulence can be regarded as a lot of small unstable air masses called eddies. It is assumed that the feature of each eddy is statistically isotropic and homogeneous. The size of such eddies ranges from an inner turbulent scale $l_0$ to an outer turbulent scale $L_0$. Furthermore, according to dimension analysis and the law of conservation of energy, it is discovered that the temperature, velocity, and refractive index of two points $i$ and $j$ satisfy the ``2/3" law that 
     \begin{equation}
         D_{TT}(L_{ij})=
         \begin{cases}
              C_T^2 L_{ij}^{2/3}, &l_0<L_{ij}<L_0\\
              C_T^2 l_0^{-4/3}L_{ij}^2, &L_{ij}\le l_0,
         \end{cases}
     \label{eq:DTT}
     \end{equation}
     \begin{equation}
         D_{nn}(L_{ij})=
         \begin{cases}
              C_n^2 L_{ij}^{2/3}, &l_0<L_{ij}<L_0\\
              C_n^2 l_0^{-4/3}L_{ij}^2, &L_{ij}\le l_0,
         \end{cases}
         \label{eq:Dnn}
     \end{equation}
     where $\langle\cdot\rangle$ represents the statistical mean. $T_i$ and $T_j$ represent the temperatures of points $i$ and $j$.  $n_i$ and $n_j$ represent the refractive indices of points $i$ and $j$, respectively. $L_{ij}$ denotes the distance between $i$ and $j$. $C_n^2$ and $C_T^2$ stand for the coefficients in~\eqref{eq:DTT} and~\eqref{eq:Dnn} in $\textrm{K}\textrm{m}^{-2/3}$ and $\textrm{m}^{-2/3}$, respectively, which are known as the refractive index structure constant (RISC) and the temperature structure index.
     \subsection{Modelling of Refractive Index Structure Constant in the Terahertz Band}
     $C_n^2$ is widely used to characterize the intensity of atmospheric turbulence. 
     Statistically, the RISC is uniform along a horizontal path, and the RISC at different altitudes $C_{n,\textrm{vis}}^2$ in the visible light frequency band with wavelength around $0.5~\mu m$ can be modelled according to the Hafnagle-Valley model~\cite{tyson1996adaptive} as
     \begin{equation}
     \begin{aligned}
             C_{n,\textrm{vis}}^2(h)&=0.00594(v/27)^2(10^{-5}h)^{10}e^{-\frac{h}{1000}}\\
             &+2.7\times 10^{-16}e^{-\frac{h}{1500}}+Ae^{-\frac{h}{100}},
     \end{aligned}
     \label{eq:RISC}
     \end{equation}
     where $h$ denotes the altitude in $\textrm{m}$, $v$ represents the average wind velocity in $\textrm{m}/\textrm{s}$, and $A$ is the terrestrial RISC in $\textrm{m}^{-2/3}$.
     However, the RISC model in the THz band has not been investigated due to the lack of measurement data. Instead, we propose a THz statistical RISC model based on the model in the visible light frequency band. This is reasonable due to the following justifications. First, based on the fact that the environmental temperature does not depend on frequency, the temperature structure constant is also frequency-invariant according to the definition~\eqref{eq:DTT}. Therefore, the temperature structure constant in the THz band is equal to that in the visible light band, i.e., $C_{T,\textrm{thz}}^2=C_{T,\textrm{vis}}^2$.
     Second, given the relationship between the refractive index and temperature in the two frequency bands $n_{\textrm{thz}}(T)$ and $n_{\textrm{vis}}(T)$, we can express the RISC model in the THz band as
     \begin{equation}
         C_{n,\textrm{thz}}^2(h)=C_{n,\textrm{vis}}^2(h)*\Big(\frac{\partial n_{\textrm{thz}}(T)}{\partial T}\Big)^2\Big/\Big(\frac{\partial n_{\textrm{vis}}(T)}{\partial T}\Big)^2.
         \label{eq:cn2_transform}
     \end{equation}
     By substituting the relationship between the refractive index versus temperature in the THz and visible light frequency bands, i.e., equations (3) and (4) in~\cite{ma2015experimental}, we can transform the RISC model in the visible light frequency band into a RISC model in the THz band according to~\eqref{eq:cn2_transform}.
    \begin{table*}[ht]  
        \caption{Statistical models of scintillation caused by turbulence.} 
        \label{table:scintillation}
        \centering
        \renewcommand\arraystretch{1}
        \begin{tabular}{p{5cm}p{5cm}p{7cm}} 
            \hline  
            \hline  
            \textbf{Name} & \textbf{Application Condition} & \textbf{Expression}\\ 
            Log-normal & $\sigma_R^2\ll 1$, Weak turbulence & $p(I)=\frac{1}{\sqrt{2\pi\sigma_{I}^2}I}\exp\left[-{(\ln I)^2}/{2\sigma_I^2}\right]$\\
            K distribution & $\sigma_R^2> 1$, Strong turbulence &
            $p(I)=\frac{2\alpha}{\Gamma(\alpha)}(\alpha I)^{(\alpha-1)/2}K_{\alpha-1}(2\sqrt{\alpha I})$
            \\
            Exponential distribution & $\sigma_R^2\gg 1$, Saturated regime & $p(I)=\frac{1}{b}\exp(-I/b)$
            \\
            Gamma-Gamma distribution & All & Equation~\eqref{eq:GammaGamma}
            \\
            \hline
            \hline  
        \end{tabular}  
    \end{table*} 
\section{Scintillation and Attenuation Modelling of Turbulence in the Terahertz Band}~\label{sec:scintillation_and_attenuation}
    Based on the characterized turbulence model with RISC, we investigate two statistical features of turbulence on THz wave propagation including scintillation and attenuation. The scintillation of turbulence characterizes the random power fluctuation of the received signal, and the attenuation of turbulence represents an additional attenuation besides the free-space path loss, molecular absorption, and scattering.  
    \subsection{Scintillation of Turbulence in the Terahertz Band}~\label{sec:scintillation}
    Unlike traditional multi-path fading led by random constructive or destructive interference of multi-path signals, turbulence scintillation is due to the random direction distortion of the LoS signal. By assuming the received signal power as $P_r$, we define the scintillation parameter $I$ as the instantaneous ratio of the signal intensity to its statistical average, i.e., $I={P_r}/{\langle P_r\rangle}$. Previous studies have proposed several statistical models to characterize the turbulence scintillation~\cite{dordova2010calculation} corresponding to the different strengths of turbulence, including the log-normal for weak turbulence, the K distribution for strong turbulence, and the exponential distribution corresponding to the saturation regime~\cite{dordova2010calculation}.
    As the criterion parameter for classification, the strength of turbulence can be distinguished according to the Rytov variance, given by 
    \begin{equation}
        \sigma_R^2=0.5C_n^2k^{7/6}L^{11/6},
    \end{equation}
    where $k=\frac{2\pi}{\lambda}$ denotes the wave number.
    The three conditions $\sigma_R^2\ll 1$, $\sigma_R^2\sim 1$, and $\sigma_R^2\gg 1$ correspond to the case of weak turbulence, strong turbulence, and saturated regime, respectively.  
    In extreme cases, e.g., when the propagation distance is $100~\textrm{km}$ and the RISC is $C_n^2=10^{-11}$, the Rytov variance at $300~\textrm{GHz}$ is as large as $\sigma_{R}^2=396$. Therefore, the strength of turbulence can experience all three conditions in the THz band, and thus it is necessary to use a universal distribution as the THz turbulence scintillation model. 
    A universal model applicable under any weak turbulence, strong turbulence or saturated regime conditions is the Gamma-Gamma distribution~\cite{al2001mathematical}. By applying Kolmogorov's theory, the Gamma-Gamma model assumes that the fluctuation of turbulence is caused by large-scale eddies and small-scale eddies, i.e., $I=I_aI_b$, and the two terms are governed by independent Gamma distributions, given by
    \begin{align}
        p_{I_a}(I_a)&=\frac{\alpha(\alpha I_a)^{\alpha-1}}{\Gamma(\alpha)}\exp(-\alpha I_a), I_a>0, \alpha>0,\\
        p_{I_b}(I_b)&=\frac{\beta(\beta I_a)^{\beta-1}}{\Gamma(\beta)}\exp(-\beta I_a), I_a>0, \beta>0,
    \end{align}
    where $\alpha$ represents the effective number of large-scale cells, and $\beta$ represents the effective number of small-scale ones.
    According to the chain rule and the independency of $I_a$ and $I_b$, the scintillation parameter $I$ follows a Gamma-Gamma distribution, which is expressed by 
    \begin{equation}
        p(I)=\frac{2(\alpha\beta)^{(\alpha+\beta)/2}}{\Gamma(\alpha)\Gamma(\beta)}I^{(\alpha+\beta)/2-1}K_{\alpha-\beta}\big[\sqrt{2(\alpha\beta I)}\big],
        \label{eq:GammaGamma}
    \end{equation}
    where $\Gamma(\cdot)$ denotes the Gamma function. $K_{p}(\cdot)$ is the modified Bessel function of the second kind of $p^{\textrm{th}}$ order. 
    The large-scale and small-scale scintillation parameters $\alpha$ and $\beta$ can be modelled according to Andrew's method~\cite{al2001mathematical}, which are expressed as
    \begin{align}
        \alpha&=\left[\exp\left(\frac{0.49\sigma_R^2}{(1+0.18D^2+0.56\sigma_R^{12/5})^{7/6}}\right)-1\right]^{-1}
        \label{eq:alpha}
        \\
        \beta&=\left[\exp\left(\frac{0.51\sigma_R^2(1+0.69D^2\sigma_R^{12/5})^{-5/6}}{(1+0.9D^2+0.62\sigma_R^{12/5})^{7/6}}\right)-1\right]^{-1}
        \label{eq:beta}
    \end{align}
    where $D=\sqrt{kl_{ra}^2/4L}$, and $l_{ra}$ represents the diameter of the antenna receiving aperture. Given that the effective area of the received antenna is $A_{\textrm{eff}}=\lambda^2/4\pi$, $l_{ra}=\lambda/\pi$. The four statistical models for turbulence scintillation are summarized in Table.~\ref{table:scintillation}. The relationship between the Gamma-Gamma distribution with the log-normal model, K distribution, and exponential distribution are elaborated as follows.
    \begin{itemize}
        \item     When $\sigma_R^2<1$, i.e., in the case of weak turbulence, we have $\alpha\gg 1$ and $\beta\gg 1$ and the Gamma-Gamma distribution is approximately a log-normal distribution. 
        \item     When $\sigma_R^2>1$, i.e., in the case of strong turbulence, we have $\beta\approx 1$, and the Gamma-Gamma distribution shrinks to a K distribution. 
        \item     When $\sigma_R^2\to \infty$, i.e., in the saturated regime turbulence, we have $\alpha\gg 1$ and $\beta\approx 1$, and the Gamma-Gamma distribution approximately follows an exponential distribution. 
    \end{itemize}

    \subsection{Attenuation Effect of Turbulence in the Terahertz Band}
    Similar to the scintillation model of turbulence, the attenuation model lacks deterministic and closed-form solutions due to the complexity and difficulty of solving the Navier-Stokes equations. An empirical formula developed by Larry C. Andrews for the turbulent attenuation $L_{\textrm{tur}}$ can be expressed as
    \begin{equation}
       L_{\textrm{tur}} =10\log\left|1-\sqrt{\sigma_{I}^2}\right|,
    \end{equation}
    where $L_{\textrm{tur}}$ denotes the attenuation caused by turbulence. $\sigma_{I}^2\triangleq\langle I^2\rangle$ denotes the variance of $I$ since $\langle I^2\rangle=1$ by definition. Given the Gamma-Gamma distribution of the scintillation parameter $I$, we have 
    \begin{equation}
    \begin{aligned}
        \sigma_I^2
        &=\langle I_a^2\rangle \langle I_b^2\rangle \\ &=\frac{1}{\alpha}+\frac{1}{\beta}+\frac{1}{\alpha\beta}.        
    \end{aligned}
    \label{eq:atten}
    \end{equation}
    By substituting~\eqref{eq:alpha} and~\eqref{eq:beta}
    into~\eqref{eq:atten}, the attenuation caused by turbulence in the THz band can be expressed in~\eqref{eq:Ltur} at the bottom of the next page.
    \setcounter{TempEqCnt}{\value{equation}}
    \begin{figure*}[b]
    \hrulefill
    \begin{equation}
    \begin{aligned}
        L_{\textrm{tur}}&=10\log_{10}\Bigg|1-\textrm{sqrt}\Bigg\{
        \exp\left(\frac{0.49\sigma_R^2}{(1+0.18D^2+0.56\sigma_R^{12/5})^{7/6}}\right)
        +\exp\left(\frac{0.51\sigma_R^2(1+0.69D^2\sigma_R^{12/5})^{-5/6}}{(1+0.9D^2+0.62\sigma_R^{12/5})^{7/6}}\right)-2\\
        &+\left[\exp\left(\frac{0.49\sigma_R^2}{(1+0.18D^2+0.56\sigma_R^{12/5})^{7/6}}\right)-1\right]\times\left[\exp\left(\frac{0.51\sigma_R^2(1+0.69D^2\sigma_R^{12/5})^{-5/6}}{(1+0.9D^2+0.62\sigma_R^{12/5})^{7/6}}\right)-1\right]
        \Bigg\}\Bigg|.
    \end{aligned}
    \label{eq:Ltur}
    \end{equation}
    \end{figure*}

\begin{figure}
		\centering
        \includegraphics[width=0.361\textwidth]{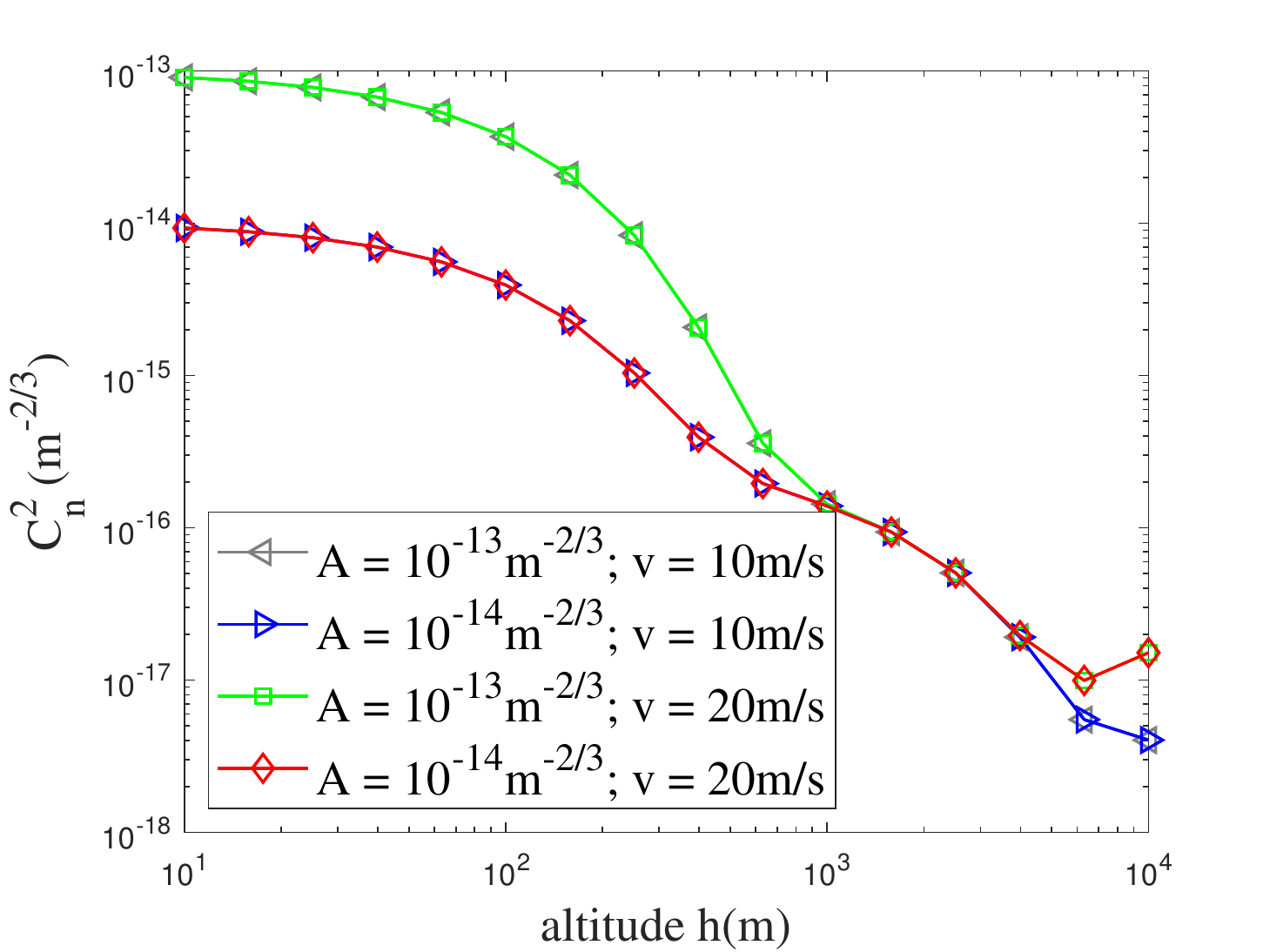}
		\captionsetup{font={footnotesize}}
        \caption{Refractive index structure constant at the different altitudes. $A$ represents the terrestrial RISC. $v$ stands for the average wind velocity.}
	    \label{fig:RISC}
	\end{figure}
    \section{Numerical Results}~\label{sec:numerical}
    In this section, we perform numerical evaluations of the effect of atmospheric turbulence on THz UAV wireless communications. Specifically, the altitude dependency of the RISC, as a critical parameter to characterize the turbulence, is first investigated. Then, the Rytov variance as an intermediate value determining the strength of turbulence at different frequencies and distances is evaluated. Furthermore, the THz scintillation and attenuation model caused by the turbulence is analyzed.
    \subsection{Refractive Index Structure Constant}
    The RISC at the different altitudes modelled in~\eqref{eq:cn2_transform} is shown in Fig.~\ref{fig:RISC} with varying terrestrial RISC $A$ and average wind speed $v$. We observe that the RISC shows a decreasing trend as the altitude increases. In the low-altitude regions ($h<1~\textrm{km}$), the terrestrial RISC $A$ primarily governs the RISC value due to the continuity of the RISC, while in the high-altitude region ($h>1~\textrm{km}$), the wind speed dominates the trend of RISC.
	
    \subsection{Rytov Variance}
\begin{figure}
		\centering
		\subfigure[]{
			\includegraphics[width=0.361\textwidth]{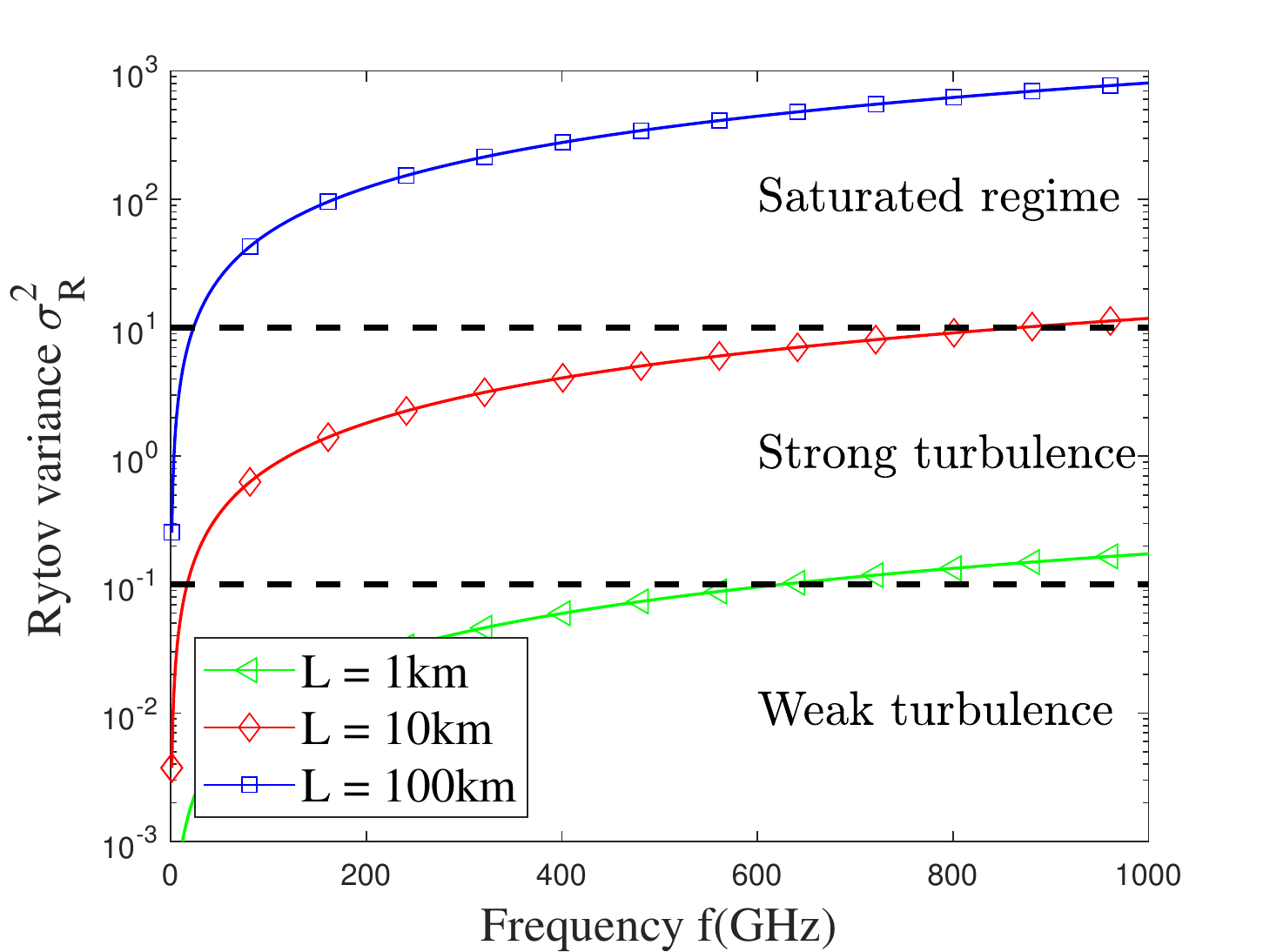}
			\label{fig:fig_rv_f_L}
		}
		\subfigure[]{
			\includegraphics[width=0.361\textwidth]{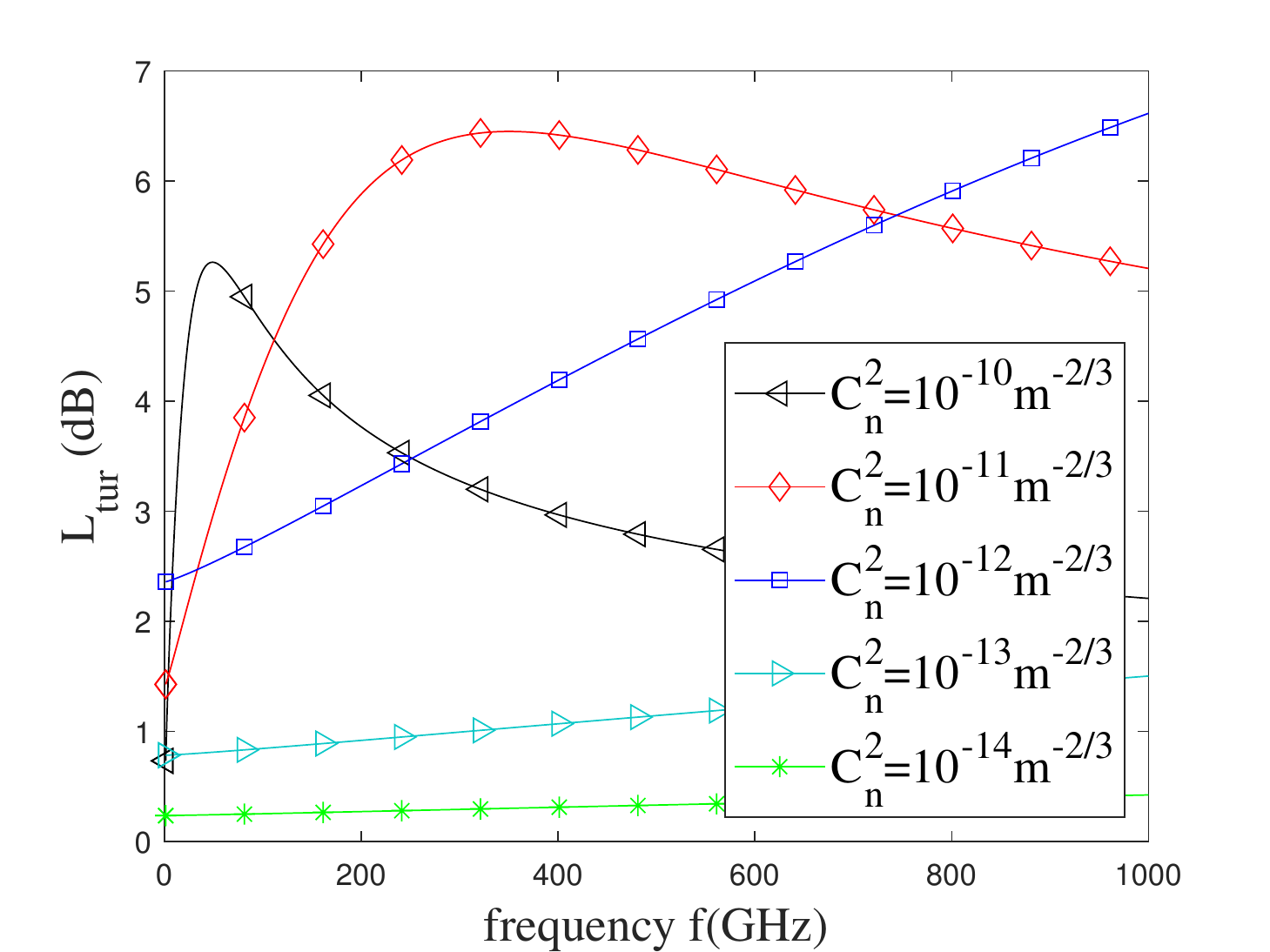}
			\label{fig:fig_rv_f_cn2}
		}
		\captionsetup{font={footnotesize}}
		\caption{Rytov variance with the different frequency. (a) With varying propagation distance $L$; (b) With varying RISC $C_n^2$. }
		\label{fig:rv_f}
	\end{figure}

    The Rytov variance determines the strength of the attenuation effect of atmospheric turbulence on THz wave propagation. The Rytov variance under different frequencies in the THz band are shown in Fig.~\ref{fig:rv_f}. In Fig.~\ref{fig:fig_rv_f_L} for varying propagation distance $L$ and RISC $C_n^2$, the RISC is $C_n^2=10^{-11}~\textrm{m}^{-2/3}$, and in Fig.~\ref{fig:fig_rv_f_cn2}, the propagation distance is $L=10~\textrm{km}$. We approximately regulate the range of weak turbulence, strong turbulence, and saturated regime as $\sigma_R^2<0.1$, $0.1\le\sigma_R^2\le 10$, and $\sigma_R^2>10$, respectively. 
    \subsection{Scintillation Caused by Atmospheric Turbulence}
\begin{figure}[h]
		\centering
        \includegraphics[width=0.361\textwidth]{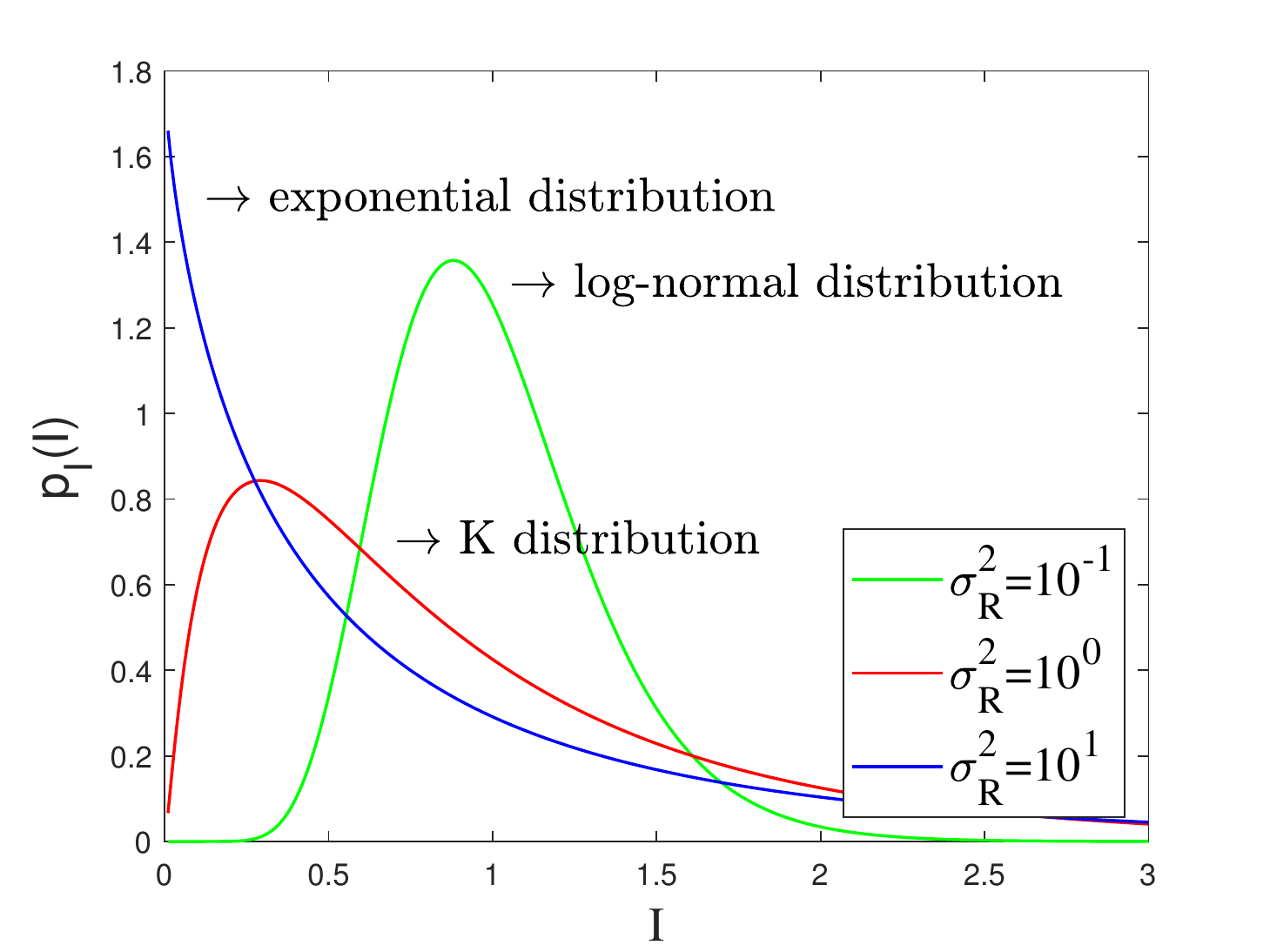}
		\captionsetup{font={footnotesize}}
        \caption{Gamma-Gamma distributed scintillation PDF versus Rytov variance.}
	    \label{fig:GG_rv}
		\vspace{-0.4cm}
	\end{figure}

    The probability density function (PDF) of the Gamma-Gamma-distributed turbulence scintillation is shown in Fig.~\ref{fig:GG_rv}. We plot the scintillation PDF with the varying Rytov variance as 0.1, 1, and 10, which corresponds to the weak turbulence, strong turbulence, and saturated regime, respectively. As we can observe, the PDFs for the three cases approximately follow log-normal, K distribution, and exponential distribution as we analyzed in Sec.~\ref{sec:scintillation}. For weak turbulence where the Rytov variance is 0.1, we have $\alpha=20.76$ and $\beta=19.75$. This indicates that both the numbers of effective large-scale and small-scale cells are large, and the log-normal distribution is reasonable due to the law of large number. When $\sigma_R^2=1$, the turbulence is strong and we have $\alpha=2.95$ and $\beta=2.46$. For the saturated regime where $\sigma_R^2=10$, we have $\alpha=2.48$ and $\beta=0.98$. 
    \subsection{Attenuation Caused by Atmospheric Turbulence}
\begin{figure}[h]
		\centering
		\subfigure[]{
			\includegraphics[width=0.361\textwidth]{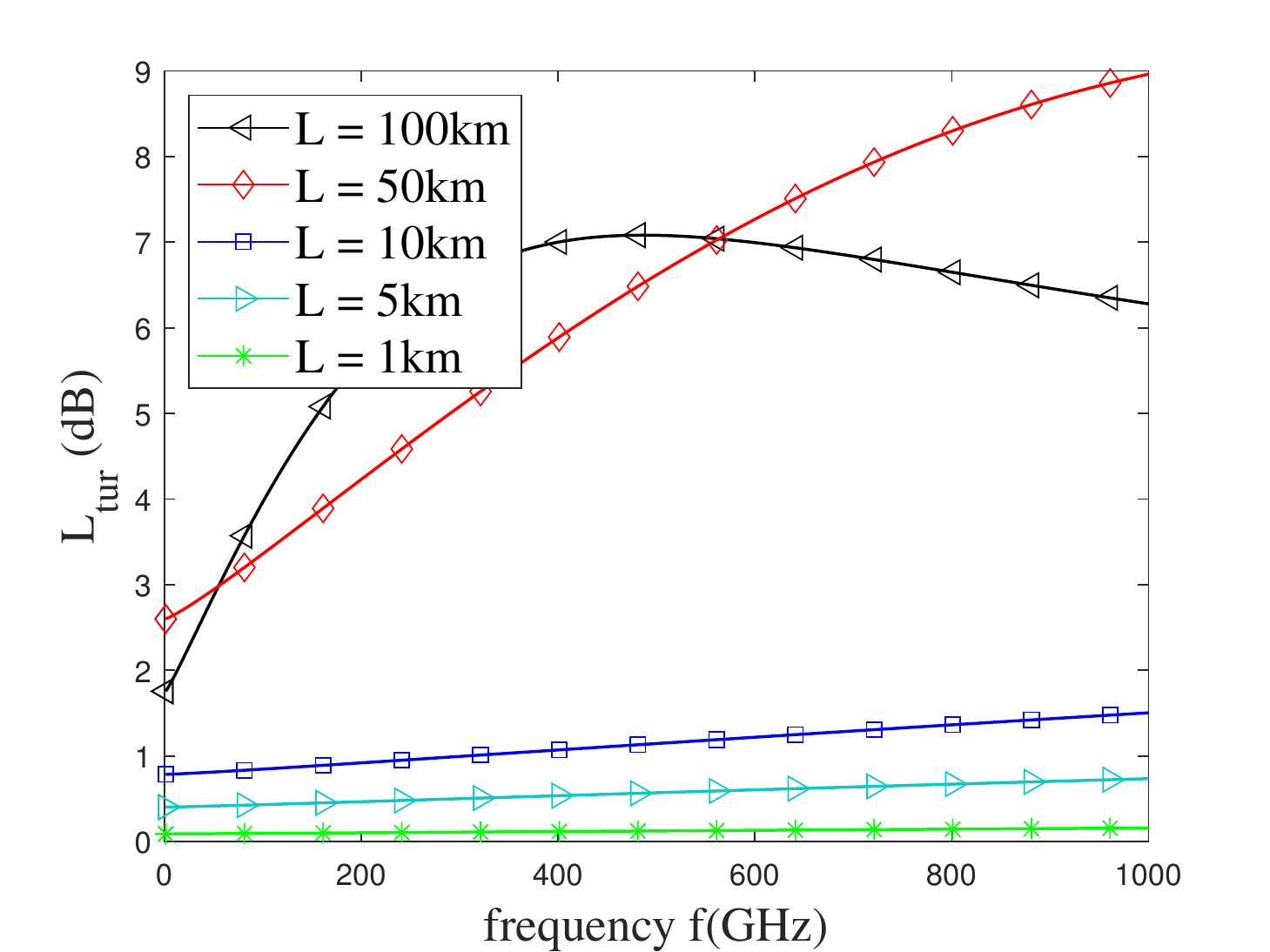}
			\label{fig:Ltur_f_L}
		}
		\subfigure[]{
			\includegraphics[width=0.361\textwidth]{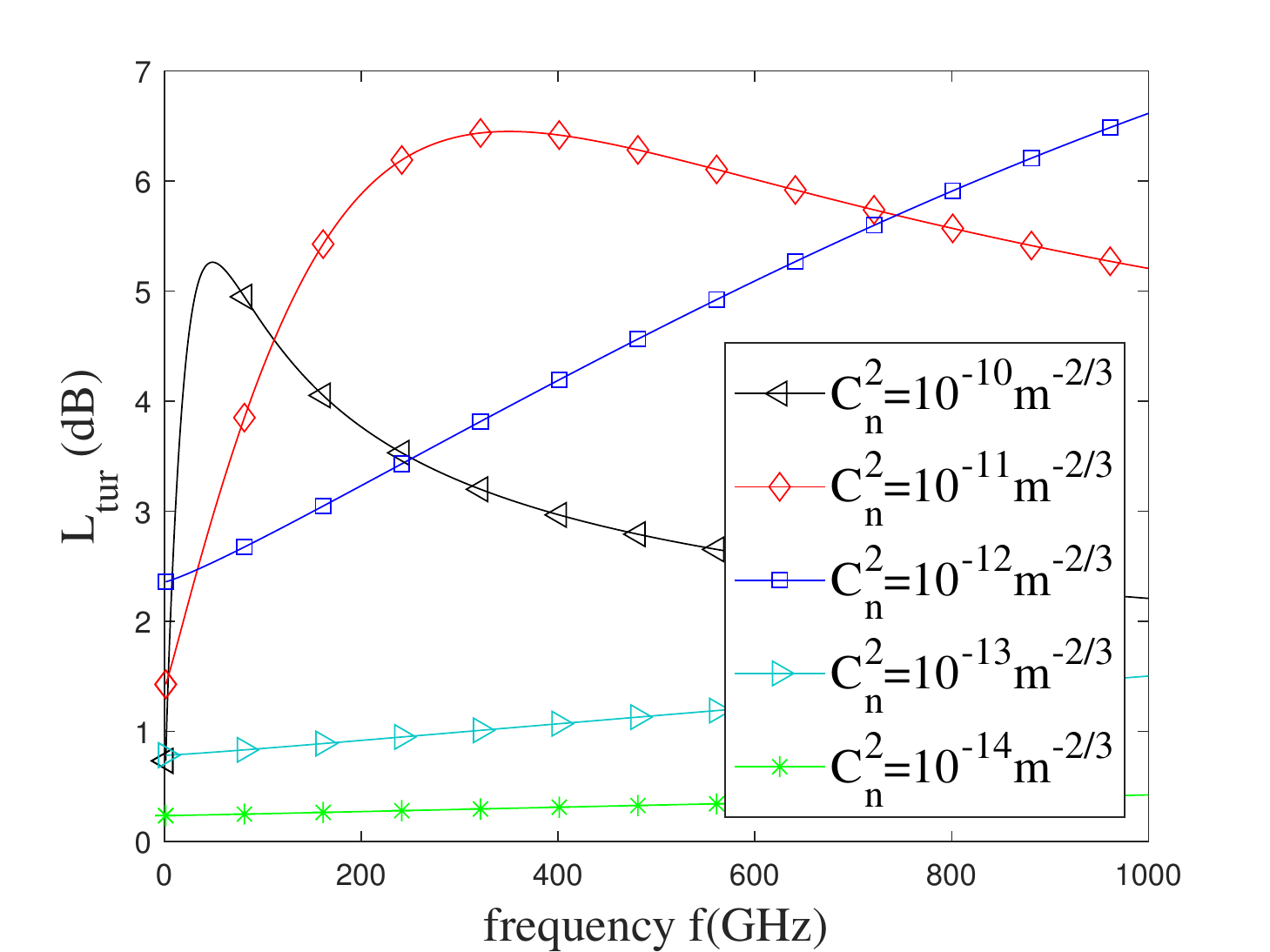}
			\label{fig:Ltur_f_cn2}
		}
		\captionsetup{font={footnotesize}}
		\caption{Attenuation caused by atmospheric turbulence with different frequencies. (a) With different propagation distance $L$; (b) With different RISC $C_n^2$. }
		\label{fig:Ltur_f}
		\captionsetup{font={footnotesize}}
		\vspace{-0.4cm}
	\end{figure}

    The attenuation caused by turbulence versus frequency with varying propagation distance $L$ and RISC $C_n^2$ is shown in Fig.~\ref{fig:Ltur_f}. In Fig.~\ref{fig:Ltur_f_L}, the RISC is taken as $10^{-13}~\textrm{m}^{-2/3}$, and in Fig.~\ref{fig:Ltur_f_cn2}, we use $L=1~\textrm{km}$. As the propagation distance and the RISC increase, the strength of the turbulence increases. Specifically, the turbulence attenuation at $1~\textrm{km}$ and $C_n^2=10^{-13}~\textrm{m}^{-2/3}$ is about $1~\textrm{dB}$. However, we observe that the increased turbulence strength does not necessarily lead to an increased turbulence attenuation. When the turbulence is weak, i.e., at a short propagation distance or low RISC, the attenuation increases with frequency, distance, and RISC. As the strength of turbulence changes from weak to strong, the attenuation of turbulence first increases and then decreases. In the THz band with frequency less than $1~\textrm{THz}$, the attenuation caused by turbulence within $10~\textrm{km}$ is less than $10~\textrm{dB}$.
	\section{Conclusion}\label{sec:concl}
	In this paper, we have investigated the THz UAV channel model in the inhomogeneous medium, and modelled the effect of atmospheric turbulence on the THz wave propagation. Specifically, the environmental parameter RISC in the THz band characterizing the intensity of turbulence is first analyzed. Then, the scintillation and attenuation characteristics caused by atmospheric turbulence are studied. The PDF of the turbulence scintillation is modelled as a Gamma-Gamma distribution, and the attenuation model based on the Gamma-Gamma scintillation is derived in a closed-form expression. Finally, numerical results demonstrate that the turbulence attenuation at $1~\textrm{km}$ and $C_n^2=10^{-13}~\textrm{m}^{-2/3}$ is approximately $1~\textrm{dB}$, which increases with the propagation distance, frequency, and RISC under the weak turbulence condition. As the strength of turbulence changes from weak to strong, the attenuation of turbulence first increases and then decreases. In the THz band with frequency less than $1~\textrm{THz}$, the attenuation caused by turbulence within $10~\textrm{km}$ is less than $10~\textrm{dB}$.
	\bibliographystyle{IEEEtran}
	\bibliography{main}
\end{document}